\documentclass[final,5p,times,twocolumn,authoryear]{elsarticle}

\usepackage{amssymb}
\usepackage{lipsum}
\usepackage{caption}
\usepackage{subcaption}
\usepackage{amsmath}

\journal{New Astronomy}

\begin{document}

\begin{frontmatter}

\title{Light Curve Models of Convective Common Envelopes}

\author[label1,label3]{N. Noughani}
\author[label2,label3]{J. Nordhaus}
\author[label1]{M. Richmond}
\author[label4]{E.C. Wilson}
\affiliation[label1]{organization={School of Physics and Astronomy, Rochester Institute of Technology},
            addressline={1 Lomb Memorial Dr.}, 
            city={Rochester},
            postcode={14623}, 
            state={NY},
            country={USA}}
\affiliation[label2]{organization={National Technical Institute for the Deaf, Rochester Institute of Technology},
            city={Rochester},
            postcode={14623}, 
            state={NY},
            country={USA}}
\affiliation[label3]{organization={Center for Computational Relativity and Gravitation, Rochester Institute of Technology}, 
            city={Rochester},
            postcode={14623}, 
            state={NY},
            country={USA}}
\affiliation[label4]{organization={Astronomy and Physics Department, Lycoming College},
            city={Williamsport},
            postcode={17701}, 
            state={PA},
            country={USA}}

\begin{abstract}
Common envelopes are thought to be the main method for producing tight binaries in the universe as the orbital period shrinks by several orders of magnitude during this phase.  Despite their importance for various evolutionary channels, direct detections are rare, and thus observational constraints on common envelope physics are often inferred from post-CE populations.  Population constraints suggest that the CE phase must be highly inefficient at using orbital energy to drive envelope ejection for low-mass systems and highly efficient for high-mass systems.  Such a dichotomy has been explained by an interplay between convection, radiation and orbital decay.  If convective transport to the surface occurs faster than the orbit decays, the CE self-regulates and radiatively cools.  Once the orbit shrinks such that convective transport is slow compared to orbital decay, a burst occurs as the release of orbital energy can be far in excess of that required to unbind the envelope.  With the anticipation of first light for the Rubin Observatory, we calculate light curve models for convective common envelopes and provide the time evolution of apparent magnitudes for the Rubin filters.  Convection imparts a distinct signature in the light curves and lengthens the timescales during which they are observable.  Given Rubin limiting magnitudes, convective CEs should be detectable out to distances of $\sim$$8$ Mpc at a rate of $\sim$$0.3$ day$^{-1}$ and provide an intriguing observational test of common envelope physics.
\end{abstract}


\end{frontmatter}

\section{Introduction}
\label{sec:intro CEs}

The primary mechanism for producing close binaries that contain at least one compact object is thought to be common envelope (CE) evolution \citep{Paczynski1976,Toonen2013,Kruckow2018,Canals2018}. CEs often occur during post-main-sequence phases, when the radius of one component of the binary expands to hundreds of times its original size, significantly increasing the interaction cross section \citep{Ivanova2012,Kochanek2014}. In two-body systems, CEs may commence via direct engulfment, Roche Lobe overflow, or orbital decay via tidal dissipation \citep{Nordhaus2010,Nordhaus2013,Chen2017}, while dynamic effects can lead to plunge in for higher-order systems \citep{Fabrycky2007,Thompson2011,Shappee2013,Michaely2016}. Two outcomes are possible: the companion survives and emerges in a post-CE binary (PCEB), or is destroyed, leaving a single star whose evolution has been significantly altered \citep{Nordhaus2011}.

The details of how energy is transferred from the orbit to the envelope is an active area of research and critical to predicting CE outcomes \citep{Macleod2018,Chamandy2018b,Wilson2019,Wilson2020,Wilson2022}. If orbital energy is released inside convective regions, turbulent eddies distribute and carry this energy toward the surface where it could be lost from the system via radiation. If this occurs before the envelope is ejected, a larger supply of energy would be required before the envelope could be unbound. Since low-mass and high-mass giant stars possess deep and vigorous convective zones, the interplay between orbital decay, convection, and radiation is necessary to understanding CE outcomes.

Despite this, current numerical simulations of CEs have not yet included convection or radiation due to computational complexity \citep{Passy2011, Ricker2012, Iaconi2019, Reichardt2020, Hatfull2021, Gonzalez2022, Chamandy2023, Ropke2023, Valsan2023}. Analytic approaches, however, can incorporate convective and radiative effects in a physically-motivated manner, and have had success reproducing the observed populations of Galactic post-CE systems for M dwarf + white dwarf (WD) systems and short-period double white dwarfs (DWDs) \citep{Wilson2019, Wilson2020}. This contends with population synthesis studies that assign constant low efficiencies to the CE phase, but still overproduce long-period binaries \citep{Politano2007, Davis2009, Zorotovic2010, Toonen2017}.

Recently, observations of high-mass, post-CE binaries in the Magellanic clouds imply that the CE phase must be highly {\it efficient} in using orbital energy to drive envelope ejection, while Galactic observations of low-mass, post-CE binaries imply that the CE phase must be highly {\it inefficient} \citep{Wilson2022}. Such a dichotomy occurs because, in high-mass systems, the time required for convection to transport orbital energy to the surface is often long compared to orbital decay, and thus the orbital energy can only contribute to envelope ejection. For low-mass CE phases, the opposite is often the case: convective transport to the surface occurs rapidly, allowing the CE to self-regulate and cool via radiative losses.

Despite the importance of the common envelope phase for determining binary evolution outcomes, direct observational tests of CEs are elusive as the evolutionary timescales are short (e.g. months to years), and the predicted transient signatures are faint \citep{Blagorodnova2016}.  With first light for the Vera C. Rubin Observatory approaching, there has been renewed interest in CE transient signatures, particularly because of their short timescales.

Motivated by the success that incorporating convection and radiation has had in reproducing Galactic post-CE populations \citep{Wilson2019, Wilson2020, Wilson2022}, we investigate how those effects impact light-curve predictions. Most noticeably, if the CE \textit{is} convective, orbital energy can be transported to the surface and lost from the system.  If this occurs faster than the orbit decays, the CE radiatively cools, and the luminosity of the system is then expected to increase \textit{gradually} while the radius of the primary remains roughly steady.  However, if the CE is {\it not} convective, the orbital energy must be used to raise the negative binding energy of the envelope.  In this case, a sudden rise in luminosity upon engulfment is expected as the envelope immediately expands \citep{Hatfull2021, Matsumoto2022}.

In this work, we hypothesize that convective eddies are able to carry energy produced by the orbiting secondary mass out of the primary star's envelope, increasing the overall brightness without contributing to unbinding the primary star's envelope. Once the orbit shrinks such that convective transport is slow compared to orbital decay, a burst occurs as the release of orbital energy can be far in excess of that required to unbind the envelope.  Since this energy can only act to drive ejection, the CE expands and more closely resembles previous predictions for CE light curves.  The lengthening of the observable timescales for convective CEs compared to non-convective CEs makes for an intriguing observational test in the era of time domain surveys.

The structure of this paper is as follows: in Section~\ref{sec:CE evolution}, we provide an overview of the CE interaction, the stellar models employed, and our assumptions. In Section~\ref{sec:CE math}, we detail how the combined effects of convection and radiation qualitatively impact CE outcomes. In Section~\ref{sec:lc calc}, we demonstrate how convection changes light-curve predictions for low-mass CE systems and give approximate observational values that can be used for real data comparison, and specific values for the Vera C. Rubin Observatory. In Section \ref{sec:analysis}, we discuss implications and estimate detection rates given Rubin limiting magnitudes. Finally, we conclude and discuss future work in Section~\ref{sec:conclusion}.

\section{Common Envelope Evolution}\label{sec:CE evolution}

The outcome of a CE phase is dependent on the physics of the interaction, the structure of the envelope, and the details of energy and angular momentum transport during the CE phase \citep{Iben1993}. A commonly-used condition for estimating whether the envelope is ejected is given by:
\begin{equation}\label{eqn:energyeff}
    E_{\rm bind} \leq \bar{\alpha}_{\rm eff} \Delta E_{\rm orb},
\end{equation}
where $E_{\rm bind}$ is the energy required to unbind the envelope, $\Delta E_{\rm orb}$ is the orbital energy released during inspiral, and $\bar{\alpha}_{\rm eff}$ is a constant, and sometimes averaged, efficiency value with which liberated orbital energy can be used to unbind the CE \citep{Tutukov1979,Iben1984,Webbink1984,Livio1988,DeMarco2011,Wilson2019, Wilson2022}. How efficiently orbital energy can be tapped to drive envelope ejection, and whether this condition is sufficient, is a subject of active research \citep{Ivanova2015, Nandez2015, Chamandy2018, Grichener2018, Ivanova2018, Soker2018, Wilson2019, Wilson2020, Wilson2022}.

Numerical simulations of CEs have had difficulty unbinding and producing post-CE orbits consistent with observations even though there is often sufficient energy available in the orbit \citep{Ricker2008,Ricker2012,Passy2011,Ohlmann2015,Chamandy2018}. This led to suggestions and debate about whether the inclusion of additional energy sources such as those from accretion, jets or recombination, as well as from processes on longer timescales could aid ejection \citep{Ivanova2015, Nandez2015, Soker2015, Kuruwita2016, Sabach2017, Glanz2018, Grichener2018, Ivanova2018, Kashi2018, Soker2018, Reichardt2020, Schreier2021, Lau2022, Valsan2023}.

Independent of energy source(s), simulations of CEs neglect important processes in post-main-sequence stars, namely the presence of vigorous and deep convective zones and the effects of radiation.  In conjunction, convection and radiation allow the CE to self-regulate and cool until orbital decay occurs faster than convective transport.  This allows the orbit to shrink substantially and ejection to occur much deeper in the envelope.  The success this approach has had in reproducing population observations of post-CE M dwarf + white dwarf binaries, double white dwarfs and massive star binaries, motivates this work \citep{Wilson2019, Wilson2020, Wilson2022}.

\subsection{Stellar Models}\label{sec:MESA models}

This work uses the open-source stellar evolution code MESA to generate spherically-symmetric models of stellar interiors (release 10108; \citealt{Paxton2011, Paxton2018}).  Models were produced for zero-age-main-sequence (ZAMS) masses ranging from $\textup{M}_\star =$ 1 \(\textup{M}_\odot\) to 6 \(\textup{M}_\odot\), where $\textup{M}_\star$ denotes the mass of the primary star in the system, in 1 \(\textup{M}_\odot\) increments from the pre-main-sequence to the white dwarf phase. The choice of initial metallicity was solar ($z=0.02$).  A Reimer's mass-loss prescription with $\eta_R=0.7$ was used for the RGB phase, while a Blöcker mass-loss prescription with $\eta_B=0.15$ was used for the 1 and 2 \(\textup{M}_\odot\) primaries, and $\eta_B=0.7$ for the 3-6 \(\textup{M}_\odot\) primaries on the AGB \citep{Reimers1975,Bloecker1995}.  These choices ensured the models matched the observed initial-final mass relationship (IFMR) derived from cluster observations \citep{Cummings2018, Hollands2023}.

For each evolutionary model, we chose the point in time when the star had evolved to its maximum radius. This yields the largest cross section for CE interactions, and makes it a reasonable time for a CE event to occur, as the primary occupies its greatest possible volume. Additionally, at this point in the evolution, the deep convective regions in the star result in strong tidal torques that act to decrease the orbits of companions that were not previously engulfed during the primary's radial expansion.  Companions that avoid engulfment, but initially orbited within $\sim$10 AU, could plunge into the envelope as tidal dissipation deposits orbital energy in the primary \citep{Villaver2009,Nordhaus2010,Nordhaus2013}.

For each primary, we selected five companions for a total of thirty possible CE interactions. We define the mass ratio of the system as ${\rm q}\equiv m_2 / M_\star$, where $m_2 < M_\star$.  For primaries with ZAMS masses between 1 and 4 $\rm M_\odot$, we chose mass ratios of  $\rm q = 0.02, 0.05, 0.08, 0.1$, and $0.2$, which correspond to companion masses between 0.02 $\rm M_\odot$ ($\sim$21 $\rm M_{\rm Jupiter}$) and 0.8 $\rm M_\odot$.  For primaries with ZAMS masses of 5 and 6 $\rm M_\odot$, mass ratios of $\textup{q} = 0.02, 0.05, 0.08, 0.1$, and $0.15$ were selected, corresponding to companion masses between 0.1 $\rm M_\odot$ and 0.9 $\rm M_\odot$. These secondaries were chosen such that convection in the primary's envelope could accommodate the luminosity produced during inspiral while remaining subsonic in nature.  In addition, each system possessed sufficient orbital energy to eject the envelope and emerge in a post-CE binary before tidal disruption occurs \citep{Guidarelli2019, Guidarelli2022}. Note that our models do not calculate any precursor emission, such as the emission that stripping the weakly bound outer layer of the star would produce \citep{Matsumoto2022}.

\section{Convective Common Envelopes}\label{sec:CE math}

The time required for the orbit of the secondary mass in a CE to fully decay, typically referred to as the inspiral timescale, is given as,
\begin{equation}\label{eqn:tinspiral}
    t_{\rm inspiral} = \int_{r_i}^{r_{\rm shred}} \frac{\left( \frac{\mathrm{d}M}{\mathrm{d}r} - \frac{M}{r}\right) \sqrt{v^2_r + (\bar{v}_{\phi}^2 + c_s^2)^2}}{4 \xi \pi G m_{2} r \rho} \,\mathrm{d}r, \
\end{equation}
where $c_s$ is the speed of sound, $G$ is the gravitational constant, $\rho$ is the gas density, $M$ is the mass interior to the orbit, $\bar{v}_{\phi}$ is the orbital velocity, $r$ is the radial position, $r_i$ is the initial radial position, and $r_{\rm shred}$ is the tidal shredding radius \citep{Nordhaus2006}. As the orbit shrinks, if the envelope is not ejected, tidal disruption of the companion will occur and result in the formation of an accretion disk inside the CE \citep{Guidarelli2022}.  This occurs where the differential gravitational force across the companion exceeds its own self gravity.  Therefore, we take the tidal shredding radius as $r_{\rm shred} \approx R_2 (2 M_{\rm core} /m_2)^{1/3}$, where $M_{\rm core}$ is the mass of the core of the primary star, $m_2$ is the mass of the companion, and $R_2$ is the radius of the companion. 

We consider three types of companions: planets, brown dwarfs and main-sequence stars and determine their radii, $R_2$, based on the type and the mass.  For $m_2 > 0.077 \rm M_\odot$, we assume the companion is a main-sequence star with radius given by 

\begin{equation}\label{eqn:companionr}
    R_{2} = \left( \frac{m_{2}}{\rm M_{\odot}}\right)^{0.92} \rm R_{\odot},\
\end{equation}

\citep{Reyes1999}.  For $0.0026 {\rm M}_\odot <m_2 < 0.077 {\rm M}_\odot$, the companion is assumed to be a brown dwarf \citep{Burrows1993} with radius given by:

\begin{equation}\label{eqn:companionr2}
    \begin{split}
     R_{2} & = {\rm R_{\odot}} \biggl[0.117 - 0.054\log^2 \left(\frac{m_{2}}{0.0026 \rm M_{\odot}}\right)\\
     & + 0.024\log^3 \left(\frac{m_{2}}{0.0026 \rm M_{\odot}}\right)\biggr] .
    \end{split}
\end{equation}

Finally, the planetary companions we consider are gas giants and assigned a radius equal to that of Jupiter \citep{Zapolsky1969}.

The orbital velocity $\bar{v}_{\phi}$ is assumed to be $\bar{v}_{\phi} = v_{\phi} - v_{\rm env} \approx v_{\phi}$, as the envelope of the primary rotates slowly compared to the orbit for RGB/AGB stars \citep{Nordhaus2007}.  Additionally, $\xi$, is a dimensionless constant that accounts for the Mach number of the companion as it moves through the envelope \citep{Park2017}. Because the motion of the companion is mildly supersonic, we take $\xi = 4$ and note that the ejection efficiency $\bar{\alpha}_{\rm eff}$ is not sensitive to this parameter for the mass ratios considered here \citep{Shima1985}.
\subsection{Effects of Convection}\label{sec:CE convection}
Red Giant and Asymptotic Giant Branch stars possess convective zones that extend from the surface deep into the interior, often making them almost fully convective.  During inspiral, energy is transferred from the orbit to the envelope.  If this energy is liberated in a convective region, it will be redistributed in the CE and transported outward.  The convective transport timescale is defined as 

\begin{equation}\label{eqn:tconv}
    t_{\rm conv} = \int_{r}^{R_\star} \frac{1}{v_{\rm conv}[r']} \,\mathrm{d}r', \
\end{equation}
and represents the time required for a convective eddy to migrate from a position in the interior to the surface.  Here, we take $v_{\rm conv}$ as the unperturbed convective velocities associated with the primary's envelope.\footnote{The convective transport timescales calculated in this work are upper limits as any acceleration of the convective eddies would result in a shorter time required to reach the surface.}

\begin{figure}
    \centering
    \includegraphics[width=0.5\textwidth]{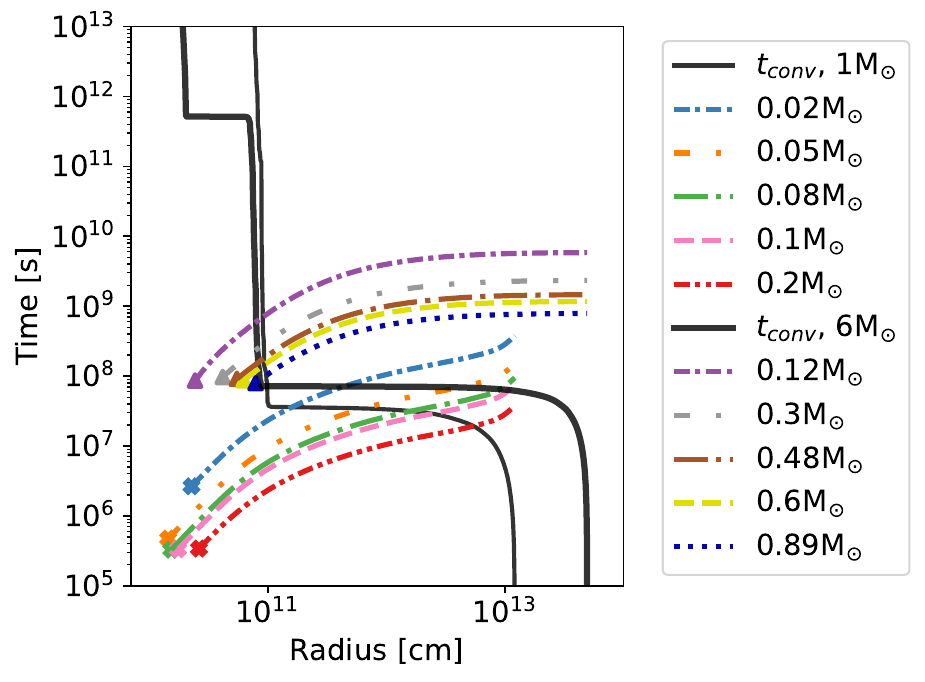}
    \caption{Above are the inspiral and convective timescales for primary stars of 1 and 6 \(\textup{M}_\odot\). The convective timescales define the time, on the y-axis, required for convection in the primary to carry energy from point $r$, on the x-axis, out to the surface ($R_\star$). These are shown as black lines for the 1 (thin) and 6 (thick) \(\textup{M}_\odot\) primaries. The inspiral timescales define the time required for the companion to inspiral from its current radius to the center of the primary star. The five companions of the 1 \(\textup{M}_\odot\) primary (0.02, 0.05, 0.08, 0.1, and 0.2 \(\textup{M}_\odot\)) and the five companion masses of the 6 \(\textup{M}_\odot\) primary (0.12, 0.3, 0.48, 0.6 and 0.89 \(\textup{M}_\odot\)) are shown above using differing lines styles and colors. The location where each companion shreds for each companion is marked with an "$\times$" (1 \(\textup{M}_\odot\) primary) or a "$\blacktriangle$" (6 \(\textup{M}_\odot\) primary).}
    \label{fig:timescales}
\end{figure}

Figure \ref{fig:timescales} compares the convective transport timescales (Eqn. \ref{eqn:tconv}) for the 1 $\rm M_\odot$ (thin-black curve) and the 6 $\rm M_\odot$ (thick-black curve) primaries to the orbital decay timescales (Eqn. \ref{eqn:tinspiral}) for ten CE configurations.  Five companions, represented by various colors and line types, are shown for each primary.  The companions associated with the 1 $\rm M_\odot$ primary represent mass ratios of $0.02, 0.05, 0.08, 0.1,$ and $0.2$, while the companions associated with the 6 $\rm M_\odot$ primary represent mass ratios of $0.02, 0.05, 0.08, 0.1,$ and $0.15$. Each curve terminates in ``$\times$" markers for the 1 $\rm M_\odot$ primary and ``$\blacktriangle$" markers for the 6 $\rm M_\odot$ primary where tidal disruption occurs. So long as the inspiral timescales (colorful curves) are greater than the convective timescales (black curves), convection can successfully carry energy out to the surface of the primary star.

As the orbit decays, the energy released is given by
\begin{equation}\label{eqn:Eorb}
    \Delta E_{\rm orb} = \frac{G m_2}{2} \left(\frac{M[r_i]}{r_i} - \frac{M[r]}{r}\right).
\end{equation}
If this energy is liberated in the convective regions defined via timescale comparisons (see Figure \ref{fig:timescales}), it can be redistributed in the CE and transported outward, eventually leaving the primary in the form of light.  The maximum luminosity that convection can accommodate while retaining its subsonic nature is given by
\begin{equation}\label{eqn:maxlum}
    L_{\rm conv, max} = \beta 4 \pi \rho r^2 c^3_{s},
\end{equation}
where we assume $\beta=5$ \citep{Grichener2018}.  As the orbit decays, the drag luminosity produced is given by
\begin{equation}\label{eqn:draglum}
    L_{\rm drag} = \xi \pi r_{\rm acc}^2 \rho v^3_{\phi},
\end{equation}
where $r_{\rm acc} = 2 G m_2 / (v^2_{\phi} + c_s)$ is the accretion radius measured from the center of the primary \citep{Nordhaus2006}.  Where the drag luminosity exceeds the maximum convective luminosity, convection may transition to the supersonic regime.  In such a case, any shocks that develop would contribute to raising the negative binding energy of the envelope.

\begin{figure}
    \centering
    \includegraphics[width=0.5\textwidth]{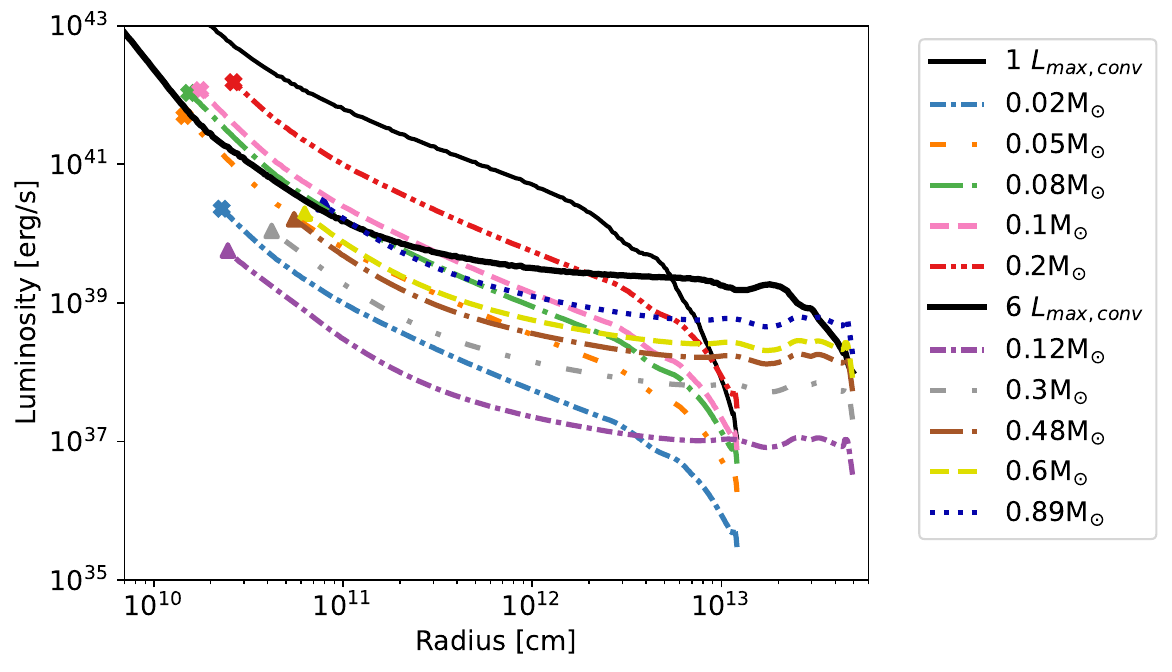}
    \caption{The maximum amount of luminosity convection can carry out of the primary stars are shown via thin (1 \(\textup{M}_\odot\)) and thick (6 \(\textup{M}_\odot\)) solid black lines labeled $L_{\rm max, conv}$. The drag luminosities produced during each CE are shown with the varying dashed and coloured lines consistent with Figure \ref{fig:timescales}. The radial limits where the companions are destroyed via tidal disruption are marked by an "$\times$"  (1 \(\textup{M}_\odot\) primary) or a "$\blacktriangle$" (6 \(\textup{M}_\odot\) primary).}
    \label{fig:draglums}
\end{figure}

Figure~\ref{fig:draglums} displays the maximum amount of luminosity convection can carry for the 1 $\rm M_\odot$ (thin-black curve) and 6 $\rm M_\odot$ (thick-black curve) primaries.  As the orbit decreases, the drag luminosity generated for each binary configuration is presented in various color and line types (consistent with Figure~\ref{fig:timescales}) with the mass of the corresponding companions listed in the legend.  The "$\times$" markers for the 1 \(\textup{M}_\odot\) primary and "$\blacktriangle$" markers for the 6 \(\textup{M}_\odot\) primary indicate where each companion is tidally disrupted. Note that convection can accommodate the totality of the luminosity generated during inspiral for the binaries presented in this work. 

The minimum energy required to unbind the primary's envelope is given as
\begin{equation}\label{eqn:Ebind}
    E_{\rm bind} = - \int_{M}^{M_{\star}} \frac{G M[r]}{r} \,\mathrm{d}m[r], \
\end{equation}
where $M_{\star}$ is the mass of the primary star. By calculating the binding energy directly from stellar evolution models, we avoid the need for $\lambda$-formalisms that approximate a star's gravitational binding energy when the interior structure is not known \citep{DeMarco2011}.

\begin{figure}
    \centering
    \includegraphics[width=0.5\textwidth]{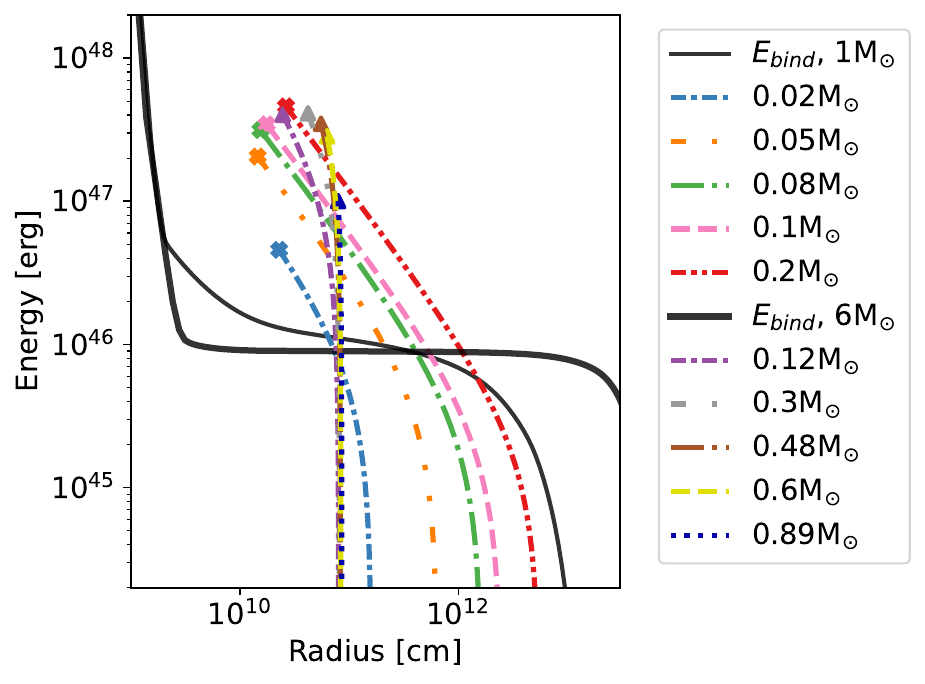}
    \caption{Illustrated above are the inspiral and binding energies for two primary masses (1 and 6 \(\textup{M}_\odot\)), with position along the radius of the primary on the x-axis in centimeters and energy in ergs on the y-axis. The binding energies for both are shown in thin (1 \(\textup{M}_\odot\)) and thick (6 \(\textup{M}_\odot\)) solid black lines, and represent the minimum energy required to unbind that portion of the primary's envelope at every point along its radius as a function of position. The companion's inspiral energies, shown in the varying colors and line patterns of Figures \ref{fig:timescales} and \ref{fig:draglums}, all intersect with the binding energy limits before reaching their shredding radii, illustrated via an "$\times$" (1 \(\textup{M}_\odot\)) or a "$\blacktriangle$" (6 \(\textup{M}_\odot\)). Therefore, all the companions shown unbind the primary's envelope and are expected to form close binary pairs.}
    \label{fig:energies}
\end{figure}

The binding energy as a function of position for the 1 $\rm M_\odot$ and 6 $\rm M_\odot$ primaries are shown in solid thin and thick black curves respectively in Figure~\ref{fig:energies}. For each primary, the energy released as the orbits decay are shown in the same color and line types as Figures~\ref{fig:timescales} and \ref{fig:draglums}.  Note where $t_{\rm conv} < t_{\rm inspiral}$, we assume the orbital energy is transported to the surface and radiated away, and therefore does not contribute to unbinding the envelope.  As the CE phase continues, orbital decay will eventually occur faster than convection can transport the energy to the surface.  This can happen while the companion still orbits in a convective zone, or will occur when it reaches a radiative zone.  In either case, the liberated orbital energy cannot escape the system and must be used to raise the negative binding energy of the envelope.  Therefore, we assume orbital energy can only contribute to unbinding the envelope when $t_{\rm inspiral} < t_{\rm conv}$.  Given these constraints, there is sufficient energy in the orbits to eject the envelope and emerge as a post-CE binary for all systems in this work (see Figure~\ref{fig:energies}).

\section{Light Curve Modeling}\label{sec:lc calc}

Our CE evolution consists of two distinct phases: (i.) a self-regulated phase and (ii.) an ejection phase.  During the self-regulated phase, orbital energy is rapidly transported to the surface and radiated away.  Since we explicitly assume this energy escapes and does not contribute to unbinding the envelope, the stellar radius remains constant while the luminosity of the system increases.  The ejection phase occurs once $t_{\rm inspiral} < t_{\rm conv}$, as the liberated orbital energy cannot be lost and must be used to drive ejection.  When this occurs, it can be the case that energy much in excess of the binding energy is supplied rapidly (see Figure~\ref{fig:energies}).

The light curve for a convective CE therefore also consists of two parts.  During the self-regulated phase, the drag luminosity drives the increase in the total luminosity of the system.  Once ejection occurs, the evolution of the luminosity is dominated by radial expansion and similar to that of Type IIP supernovae, which are characterized by a plateau phase \citep{Chugai1991, Popov1993, Kasen2009}.  Note that the application of Type IIP supernova light curve modeling to low-mass, non-convective CE and merger events has been previously studied in detail \citep{Ivanova2013, Hatfull2021, Matsumoto2022,Chen2024}.

\subsection{Luminosity Evolution During Ejection}

For the ejection phase, we employ a simple model where we assume that the gas is optically thick at temperatures above a chosen ionization temperature, $T_{\rm ion}$, and optically thin at lower temperatures \citep{Popov1993}.  This results in a two-zone model in the envelope with a recombination front that starts at time $t_i$ and initial position $r=R_{i}$ serving as the boundary.  We define an envelope expansion timescale, $t_{\rm exp}\equiv R_\star/v_{\rm exp} = R_\star\left(3M/10E\right)^{1/2}$, where $R_\star$ is the radius of the primary star, $v_{\rm exp}$ is the expansion velocity, $M$ is the mass of the ejected envelope, and $E$ is the energy of the ejected envelope, which we take to be equal to the orbital energy that is released at the location where the envelope is ejected (Figure \ref{fig:energies}).  In combination with the photon diffusion timescale, $t_{d}=9 \kappa M/4 \pi^{3} c R_\star$, we define a characteristic timescale on which changes occur as $t_a\equiv \left(2t_dt_{\rm exp}\right)^{1/2}$.  We choose an ionization temperature of 5000 K \citep{Popov1993} for all models, and set the opacity to the corresponding value at that temperature in the primary's interior.  This results in typical expansion speeds ranging from $\sim$50 to 100 km/s, consistent with outflow ejecta in proto-planetary and planetary nebulae \citep{Bujarrabal2001, Lorenzo2021}.

Recombination begins at moment $t_i$, which is determined by equating the luminosity on the interior-side and exterior-side of the recombination front such that

\begin{equation}\label{eqn:ti}
    \frac{E}{2t_d} e^{- t_i ^2/t_a ^2} = 8 \pi \sigma_{\rm SB} v_{\rm exp} ^2 t_i ^2 T_{\rm ion} ^4,
\end{equation}
where $\sigma_{\rm SB}$ is the Stefan-Boltzmann constant \citep{Popov1993}.  Note that $t_i$ is measured from the moment expansion begins, i.e. the moment orbital decay occurs faster than convective transport. The time evolution of the luminosity during the plateau phase is then given by

\begin{equation}\label{eqn:lum_sn}
    L_{\rm bol} = 8 \pi \sigma_{\rm SB} T_{\rm ion} ^4 v_{\rm exp} ^2 \left[t_i t \left(1 + \frac{t_i ^2}{3 t_a ^2}\right) - \frac{t^4}{3 t_a ^2}\right].
\end{equation}

During the ejection phase, the envelope expands and cools, and formation of dust may initially obscure the natal white dwarf \citep{Bermudez2024}.  Observations of post-RGB/AGB systems often show a breaking of symmetry with dusty tori in the orbital plane accompanied by bipolar outflows or jets perpendicular to the orbital plane \citep{Winckel2003}.  Such nebular shaping is an expected outcome of CE evolution and consistent with systems in which the linear momentum and energy in the outflows far surpass what single stars can achieve \citep{Bujarrabal2001,Nordhaus2007}.  Thus, the evolution of the post-plateau luminosity depends on orientation and whether the hot, natal white dwarf can be observed, perhaps through cavities in the ejecta caused by jets clearing material.  We calculate upper limits for the luminosity assuming that the white dwarf core is visible after the envelope expands for one year.  Given our range of expansion velocities, this typically occurs when the envelope reaches ten to twenty astronomical units.

\begin{figure*}%
    \centering
    \includegraphics[width=1.0\textwidth]{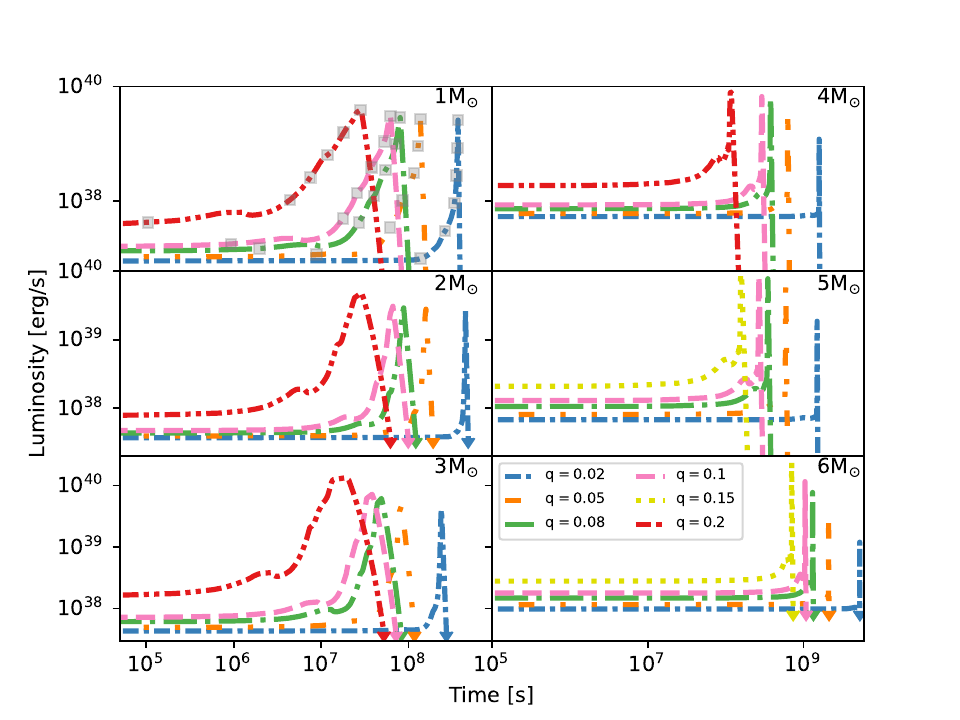}
    \caption{Time evolution of the predicted light curves for convective common envelopes are shown from top to bottom, then left to right, for primary masses of 1, 2, 3, 4, 5, and 6 ${\rm M}_\odot$.  For each primary, five CE light curves for different mass ratios are displayed with various colors and line styles.  The 1 \(\textup{M}_\odot\) primary plot includes grey boxes representing times where apparent magnitudes were calculated (see Section \ref{sec:obs approx} for more detail).  For each CE, we choose six of these times by uniformly sampling the corresponding light curve between its maximum value and 1.1 times its minimum value in log-space.  The seventh point is taken after the envelope is ejected and represents an upper limit on the magnitude if the natal white dwarf were visible one year post-CE expulsion (depicted by upper limit triangle symbols above).}
    \label{fig:lightcurves}
\end{figure*}

Figure \ref{fig:lightcurves} shows the full time evolution of the light curves for all thirty binary configurations, with panels representing primary mass increasing top to bottom and then left to right.  The colors of the curves and their line styles are consistent with the previous figures in this paper, but are shown in terms of the corresponding mass ratios.  The 1 \(\textup{M}_\odot\) primary star plot also includes grey boxes, which denote all the points (but one, the final WD luminosity) where an apparent magnitude was calculated for the models, as described in the following section. 

We note that for all models, the evolution of the light curve is dominated by the self-regulated stage rather than the ejection phase. Convection extends the time required until envelope ejection occurs as energy is lost from the system.   The imprint of the drag luminosity becomes the defining characteristic of the light curves for these events, and provides a means of identification of a CE in progress, pre-envelope ejection. 

\subsection{Expected Observational Magnitudes}\label{sec:obs approx}
\label{sec:mags}

The Vera C. Rubin Observatory's Simonyi Survey telescope \citep{LSST2019} will measure targets in AB magnitude; therefore, we make our anticipated magnitude calculations in this system.  Apparent magnitudes were derived for each model by applying a blackbody approximation over a broad wavelength range. We calculated a flux by integrating over the filter transmission curves. These curves were normalized for integration for each of the telescope's filters across their respective wavelength ranges \citep{Kahn2010}. We calculated several properties for each filter: a central wavelength, $\lambda_c$, equivalent width in wavelength, $\Delta_\lambda$, and equivalent width in frequency, $\Delta_\nu$. We integrated
the flux density of each model over each respective equivalent wavelength range to derive an overall flux, $Q$, in each of the telescope's passbands.

We then divided this flux by the filter width $\Delta_\nu$, to convert this flux into a flux density per unit frequency, $\Phi_\nu=Q/\Delta_\nu$.  We use $\Phi_\nu$ to calculate the AB magnitude:
\begin{equation}\label{eqn:mag}
    m = -2.5 \log\left(\frac{\Phi_\nu}{\Phi_0}\right),
\end{equation}
 where $\Phi_0$ is the zero point flux density for the AB magnitude system. We calculated these magnitudes at varying distances by scaling our flux density as $f\Phi_\nu$, where $f=R_p^2/D^2$, $R_p$ is the radial size of the photosphere, and $D$ is the distance to the system.

\begin{figure*}
  \begin{subfigure}[t]{.49\textwidth}
    \centering
    \includegraphics[width=\linewidth]{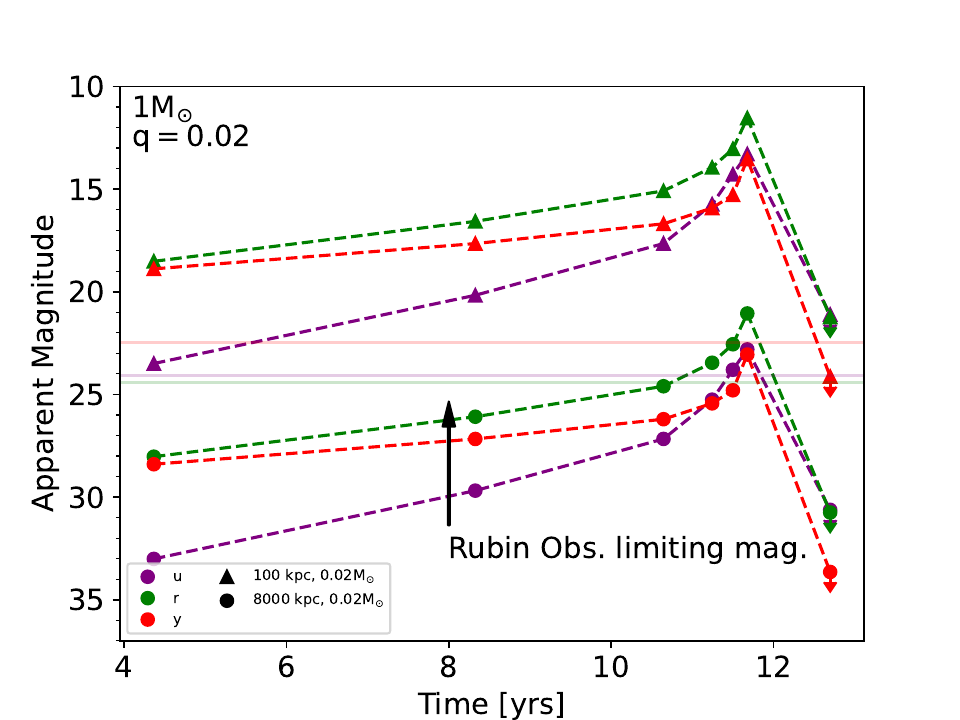}
  \end{subfigure}
  \hfill
  \begin{subfigure}[t]{.49\textwidth}
    \centering
    \includegraphics[width=\linewidth]{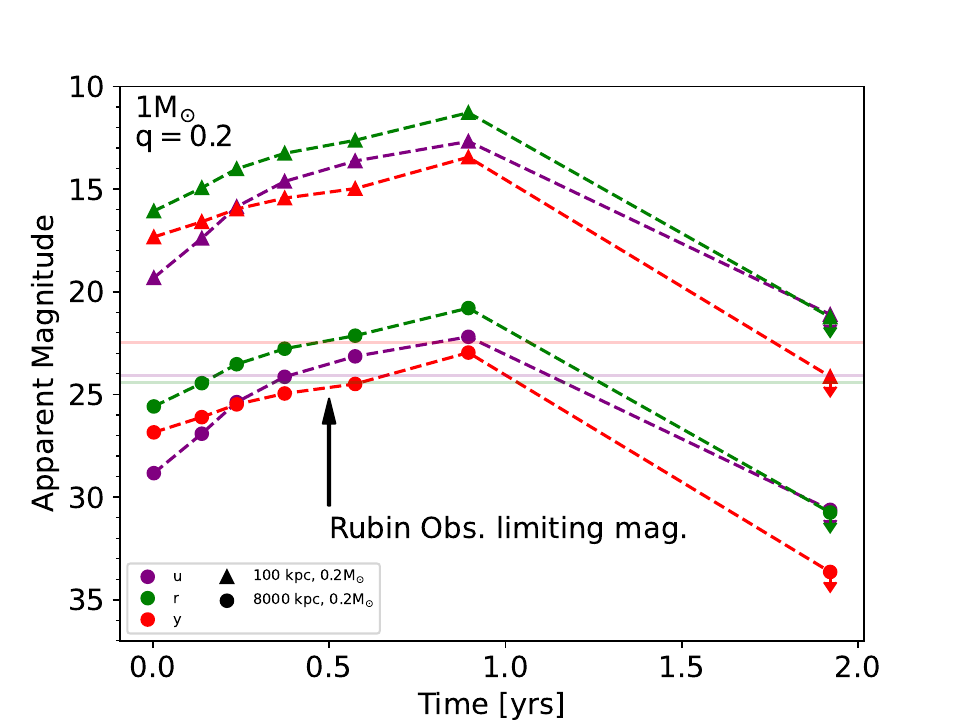}
  \end{subfigure}

  \smallskip

  \begin{subfigure}[t]{.49\textwidth}
    \centering
    \includegraphics[width=\linewidth]{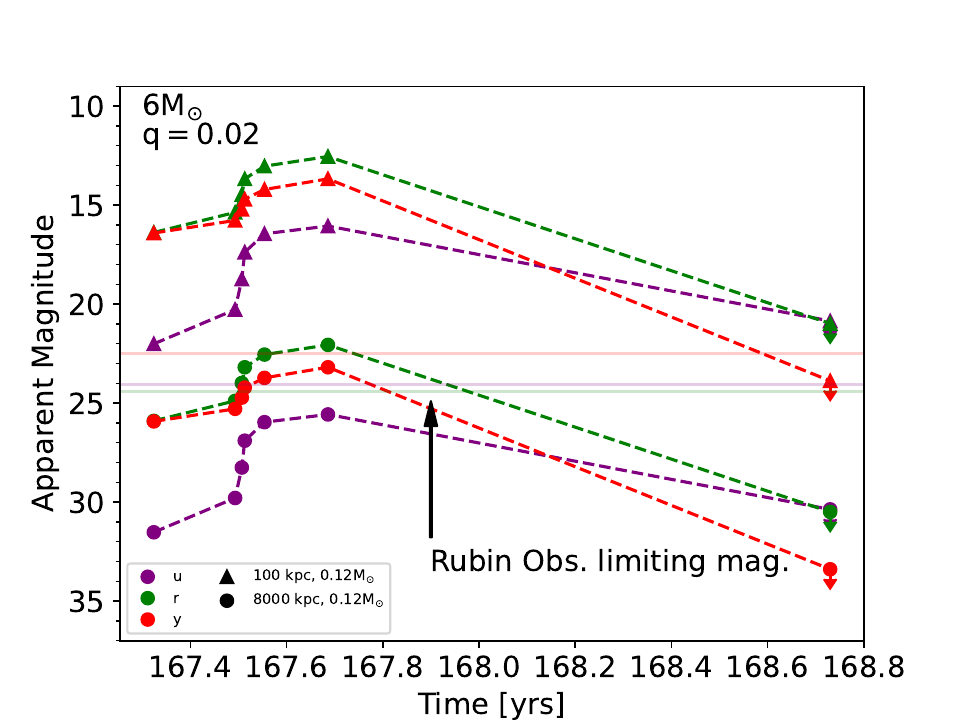}
  \end{subfigure}
  \hfill
  \begin{subfigure}[t]{.49\textwidth}
    \centering
    \includegraphics[width=\linewidth]{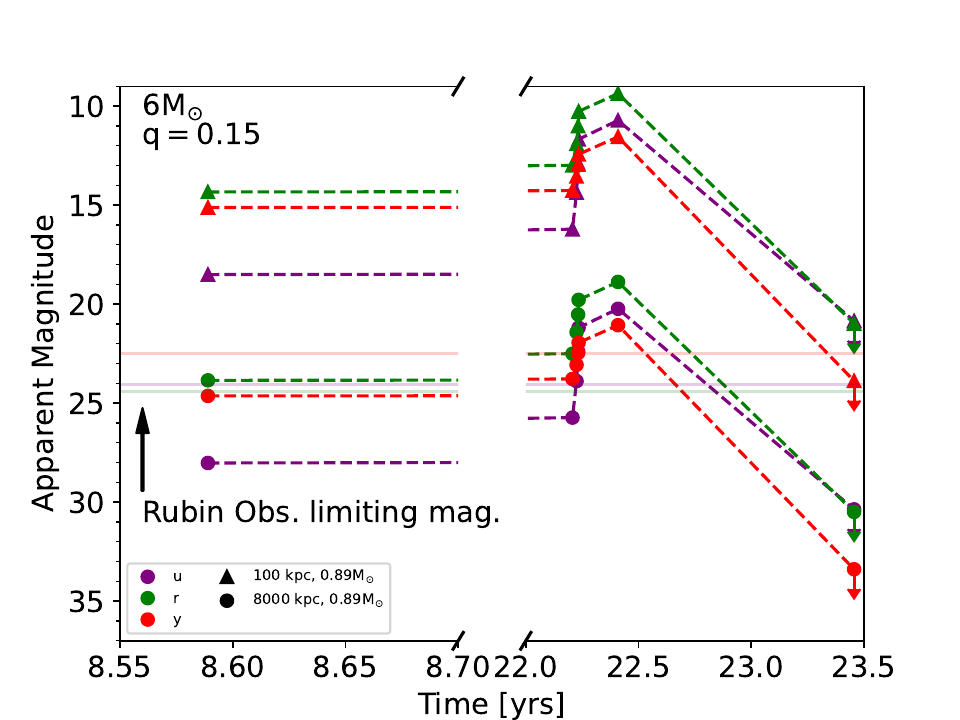}
  \end{subfigure}
  \caption{Apparent AB magnitudes for the 1 \(\textup{M}_\odot\) primary with the smallest mass ratio system (${\rm q}=0.02$; top left) and the largest mass ratio system (${\rm q}=0.2$; top right) and the 6 \(\textup{M}_\odot\) primary with the smallest mass ratio system (${\rm q}=0.02$; bottom left) and the largest mass ratio system (${\rm q}=0.15$; bottom right). Each plot shows three of the six Rubin filters, namely, u, r, and y. The semi-transparent, horizontal lines on each plot display the limiting magnitudes for the instrument, where the colors are consistent with each filter. Magnitudes are shown for two distances: 100 kpc (triangles) and 8000 kpc (circles). The points illustrated are those sampled from various epochs in the light curves as described in Section~\ref{sec:obs approx} and shown in Fig. \ref{fig:lightcurves}.  The last evolutionary point represents an upper limit assuming the natal white dwarf is visible.}
  \label{fig:appmags}
\end{figure*}

In Figure~\ref{fig:appmags}, we present apparent magnitudes corresponding to seven discrete times during CE evolution.  For each binary, we choose six of these times by uniformly sampling the corresponding light curve between its maximum value and 1.1 times its minimum value (see gray boxes in top-left panel of Figure~\ref{fig:lightcurves}).  The seventh point is taken after the envelope is ejected and represents the earliest time the natal white dwarf may be visible.  All panels in Figure~\ref{fig:appmags} depict three of the six Rubin Observatory bands -- u-band (purple symbols), r-band (green symbols) and y-band (red symbols) -- at two distances, 100 kiloparsecs (triangles) and 8000 kiloparsecs (circles).  The limiting magnitudes for the telescope are also shown on each plot in horizontal lines with colors corresponding to their respective filters \citep{Kahn2010}.  Apparent magnitudes that are dimmer than the corresponding filter's limiting magnitude will not be detectable by the instrument in that band.  

Apparent magnitudes of 30 common envelope systems were calculated and compared to the Rubin filter limiting magnitudes, and a representative sample are shown in Figure~\ref{fig:appmags}, again shown at the seven discrete times from Figure \ref{fig:lightcurves}.  Tables~\ref{table:Table1}-\ref{table:Table6} list absolute magnitudes in all six instrument bands (u, g, r, i, z, y) for the lowest mass-ratio system for each primary star, also at the seven discrete times described in Figure \ref{fig:lightcurves}. Note that these systems produce the faintest predicted signatures while still being detectable, thereby implying that higher-mass ratio systems at the same distance will also be observable.  The last line of each table provides a magnitude consistent with the upper luminosity limit if the hot, natal white dwarf is visible to the observer. 

\begin{table}
	\resizebox{\columnwidth}{!}{\begin{tabular}{l c c c c c r}
		\hline
		\textbf{Time [yrs]} & \textbf{u} & \textbf{g} & \textbf{r} & \textbf{i} & \textbf{z} & \textbf{y}\\
		\hline
		4.36 & 3.50& 0.219& -1.48& -2.34& -2.69& -1.11\\
		8.32 & 0.172& -2.33& -3.43& -3.96& -4.10& -2.35\\
		10.63 & -2.35& -4.27& -4.91& -5.20& -5.19& -3.31\\
        11.23 & -4.26& -5.75& -6.06& -6.17& -6.05& -4.08\\
        11.49 & -5.72& -6.89& -6.96& -6.95& -6.75& -4.72\\
        11.66 & -6.71& -8.17& -8.46& -8.56& -8.44& -6.46\\
        12.68& 1.11*& -0.708*& 1.24*& 1.55*& 1.93*& 4.13*\\
		\hline
	\end{tabular}}
    \caption{The table below shows the absolute magnitude for the 1 \(\textup{M}_\odot\) primary and 0.02 \(\textup{M}_\odot\) companion. The times chosen are those marked by the squares in Figure \ref{fig:lightcurves} and are measured from the start of the CE phase. The symbol * denotes when we've calculated upper limits to the emission.}
 \label{table:Table1}
\end{table}

\begin{table}
	\resizebox{\columnwidth}{!}{\begin{tabular}{l c c c c c r}
		\hline
		\textbf{Time [yrs]} & \textbf{u} & \textbf{g} & \textbf{r} & \textbf{i} & \textbf{z} & \textbf{y}\\
		\hline
		10.36& 3.11& -0.547& -2.54& -3.57& -4.02& -2.53\\
		13.35& 0.228& -2.76& -4.23& -4.97& -5.23& -3.59\\
		13.75& -2.11& -4.56& -5.60& -6.11& -6.23& -4.46\\
        13.94& -4.02& -6.02& -6.72& -7.05& -7.05& -5.19\\
        14.03& -5.56& -7.21& -7.64& -7.82& -7.74& -5.80\\
        14.21& -6.10& -7.97& -8.56& -8.83& -8.80& -6.90\\
        15.26& -1.16*& -1.56*& -1.03*& -0.717*& -0.332*& 1.87*\\
		\hline
	\end{tabular} }
    \caption{The table below shows the absolute magnitude for the 2 \(\textup{M}_\odot\) primary and 0.04 \(\textup{M}_\odot\) companion. The symbol * denotes when we've calculated upper limits to the emission.}
 \label{table:Table2}
\end{table}

\begin{table}
	\resizebox{\columnwidth}{!}{\begin{tabular}{l c c c c c r}
		\hline
		\textbf{Time [yrs]} & \textbf{u} & \textbf{g} & \textbf{r} & \textbf{i} & \textbf{z} & \textbf{y}\\
		\hline
		4.59 & 2.53& -0.99& -2.88& -3.85& -4.26& -2.75\\
		6.56 & -0.406& -3.24& -4.60& -5.28& -5.50& -3.83\\
		7.06 & -2.76& -5.05& -5.98& -6.43& -6.51& -4.71\\
        7.23 & -4.65& -6.51& -7.10& -7.36& -7.33& -5.44\\
        7.33 & -6.16& -7.68& -8.00& -8.13& -8.02& -6.05\\
        7.53 & -6.78& -8.53& -9.03& -9.25& -9.19& -7.27\\
        8.59& -1.20*& -1.60*& -1.07*& -0.756*& -0.372*& 1.83*\\
		\hline
	\end{tabular}}
 \caption{The table below shows the absolute magnitude for the 3 \(\textup{M}_\odot\) primary and 0.06 \(\textup{M}_\odot\) companion. The symbol * denotes when we've calculated upper limits to the emission.}
  \label{table:Table3}
\end{table}

\begin{table}
	\resizebox{\columnwidth}{!}{\begin{tabular}{l c c c c c r}
		\hline
		\textbf{Time [yrs]} & \textbf{u} & \textbf{g} & \textbf{r} & \textbf{i} & \textbf{z} & \textbf{y}\\
		\hline
		47.08 & 2.90& -0.820& -2.86& -3.92& -4.38& -2.91\\
		50.19 & 0.715& -2.49& -4.14& -4.98& -5.30& -3.71\\
		50.29 & -1.16& -3.93& -5.24& -5.89& -6.10& -4.40\\
        50.33 & -2.78& -5.18& -6.19& -6.68& -6.79& -5.01\\
        50.35 & -4.17& -6.25& -7.01& -7.36& -7.39& -5.54\\
        50.49 & -4.50& -6.72& -7.60& -8.02& -8.08& -6.26\\
        51.52& 1.39*& 0.995*& 1.52*& 1.84*& 2.22*& 4.42*\\
		\hline
	\end{tabular}}
 \caption{The table below shows the absolute magnitude for the 4 \(\textup{M}_\odot\) primary and 0.08 \(\textup{M}_\odot\) companion. The symbol * denotes when we've calculated upper limits to the emission.}
  \label{table:Table4}
\end{table}

\begin{table}
	\resizebox{\columnwidth}{!}{\begin{tabular}{l c c c c c r}
		\hline
		\textbf{Time [yrs]} & \textbf{u} & \textbf{g} & \textbf{r} & \textbf{i} & \textbf{z} & \textbf{y}\\
		\hline
		43.65 & 2.32& -1.29& -3.25& -4.26& -4.69& -3.20\\
		47.28 & 0.071& -3.02& -4.57& -5.35& -5.64& -4.03\\
		47.35 & -1.85& -4.49& -5.69& -6.29& -6.46& -4.74\\
        47.38 & -3.48& -5.74& -6.65& -7.08& -7.16& -5.35\\
        47.39 & -4.87& -6.82& -7.47& -7.77& -7.76& -5.89\\
        47.54 & -5.31& -7.38& -8.14& -8.49& -8.51& -6.66\\
        48.58& 1.15*& 0.748*& 1.28*& 1.59*& 1.97*& 4.17*\\
		\hline
	\end{tabular}}
    \caption{The table below shows the absolute magnitude for the 5 \(\textup{M}_\odot\) primary and 0.1 \(\textup{M}_\odot\) companion. The symbol * denotes when we've calculated upper limits to the emission.}
  \label{table:Table5}
\end{table}

\begin{table}
	\resizebox{\columnwidth}{!}{\begin{tabular}{l c c c c c r}
		\hline
		\textbf{Time [yrs]} & \textbf{u} & \textbf{g} & \textbf{r} & \textbf{i} & \textbf{z} & \textbf{y}\\
		\hline
		167.11 & 2.00& -1.64& -3.62& -4.64& -5.08& -3.60\\
		167.28 & 0.272& -2.97& -4.64& -5.48& -5.81& -4.23\\
		167.29 & -1.27& -4.15& -5.54& -6.23& -6.46& -4.80\\
        167.30 & -2.62& -5.19& -6.33& -6.89& -7.04& -5.30\\
        167.34 & -3.56& -5.96& -6.97& -7.46& -7.57& -5.79\\
        167.47 & -3.95& -6.40& -7.46& -7.97& -8.10& -6.33\\
        168.52& 0.847*& 0.448*& 0.976*& 1.29*& 1.67*& 3.87*\\
		\hline
	\end{tabular}}
    \caption{The table below shows the absolute magnitude for the 6 \(\textup{M}_\odot\) primary and 0.12 \(\textup{M}_\odot\) companion. The symbol * denotes when we've calculated upper limits to the emission.}
 \label{table:Table6}
\end{table}

\begin{figure*}%
    \centering
    \includegraphics[width=1.0\textwidth]{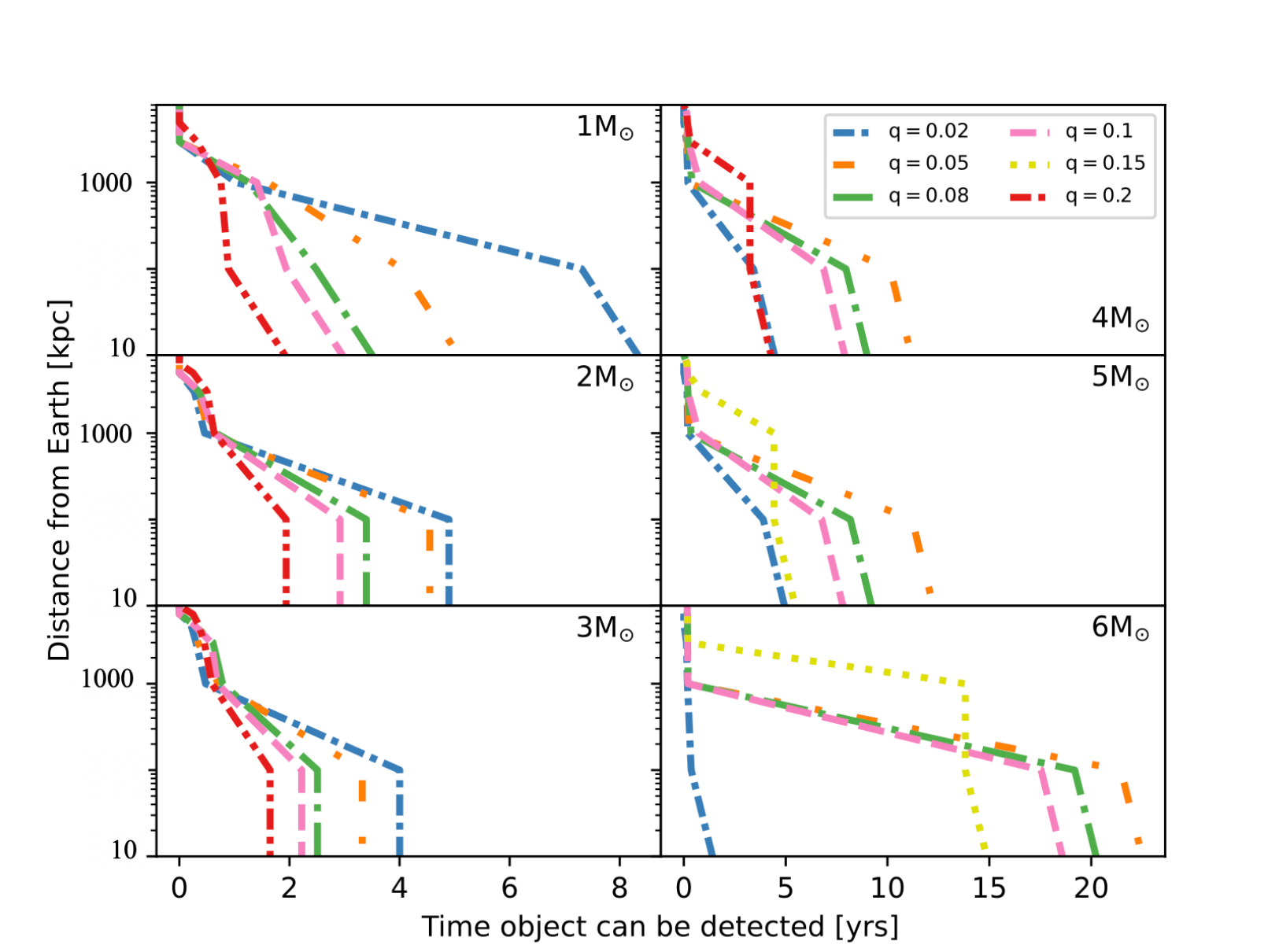}
    \caption{Above from top to bottom, then left to right, are plots demonstrating how long these convective CE systems are expected to be observed by the Rubin Observatory instrument, with time along the x-axis, for distances up to 8000 kpc, shown along the y-axis, for primary masses of 1, 2, 3, 4, 5, and 6 \(\textup{M}_\odot\) with their respective five companion masses in various colors and line styles. The companions represent mass ratios of 0.02, 0.05, 0.08, 0.1 and 0.2 (primaries 1-4 \(\textup{M}_\odot\)), and 0.02, 0.05, 0.08, 0.1, and 0.15 (primaries 5 and 6 \(\textup{M}_\odot\)). As the LSST dataset is anticipated to span 10 years, many of the CE systems shown are ideal candidates and have a high likelihood of being observed by the telescope.}
    \label{fig:howlongall}
\end{figure*}

\section{Results and Analysis}\label{sec:analysis}

The Rubin Observatory will observe the southern sky every few nights for a decade \citep{LSST2019}.  Given the limiting magnitudes for the Rubin Observatory instrument, we calculate the total time during which a common envelope event is visible and present the results in Figure~\ref{fig:howlongall}.  The ordinate shows the distance to the CE event from the Earth while the abscissa presents the total time the CE can be observed, measured in years.  Color and line styles are consistent with previous figures.  Note that we define visibility such that the CE event is detectable in all filters at all discrete time points in its evolution.  Thus, visibility is a measure of the total time during which the Rubin Observatory could observe a common envelope event.

Figure \ref{fig:howlongall} shows that for a large combination of primary masses and mass ratios, common envelopes can be detected by Rubin for galactic and extra-galactic sources.  As the primary mass increases, the differences between convective CEs and non-convective CEs become more prominent. For example, for a 1$\rm{M}_\odot$ primary, the observable timescale increases by 22 days for the highest mass ratio system ($\rm{q}=0.2$) and by 146 days for the lowest mass ratio system ($\rm{q}=0.02$) compared to the non-convective case. For a 6$\rm{M}_\odot$ primary, the observable timescale lengthens by 21 years for a system with $\rm{q}=0.15$ and 124 years for a system with $\rm{q}=0.02$.  Given the duration of Rubin operations, distinguishing between convective and non-convective CEs should be possible.

To estimate the number of CE events that Rubin might observe, we first calculate an approximate number of CE events occurring at any point in time in the Milky Way as:

\begin{equation}\label{eqn:CE rate}
    {N}_{\rm MW} \sim f_{\rm CE} n_{\rm s, MW} t_{\rm CE} / \bar{t},
\end{equation}

where $N_{\rm MW}$ is the number of CEs occurring in the Milky Way, $\bar{t}$ is the average lifetime of a typical star, $t_{\rm CE}$ is the timescale for a typical CE phase, $n_{\rm s, MW}$ is the number of stars in the Milky Way, and $f_{\rm CE}$ is the fraction of these stars which could incur a CE.  Taking $f_{\rm CE}=0.2$, $n_{\rm s, MW}=10^{11}$, $\bar{t}=10^{10}$ years, and $t_{\rm CE}=1$ year, yields ${N}_{\rm MW} \sim 2$ CEs occurring in the Milky Way.  Extending this number to the Local Group can be approximated by ${N}_{\rm LG} \sim \left(\Upsilon_{\rm LG}/\Upsilon_{\rm MW}\right) {N}_{\rm MW}$ where $\Upsilon_{\rm LG}$ is the mass-to-light ratio in the Local Group and $\Upsilon_{\rm MW}$ is the mass-to-light ratio in the Milky Way.  Adopting $\Upsilon_{\rm LG}\sim 2 \Upsilon_{\rm MW}$ yields ${N}_{\rm LG} \sim 4$ \citep{Bergh1999,Overduin2004}.  Given that Rubin could detect CEs to $\sim$$8$ Mpc, we calculate an overall conservative CE number as $N_{\rm 8Mpc} \sim \left(\Upsilon_{\rm 8Mpc}/\Upsilon_{\rm MW}\right) {N}_{\rm MW} \sim 130$ \citep{Karachentsev2004} which would correspond to an approximate rate of one CE every three days.

\section{Conclusions}\label{sec:conclusion}

In this work, we present observational signatures of convective common envelopes and demonstrate that such events will be visible by the Rubin Observatory's Simonyi Survey Telescope. Following the dynamics initially described in \cite{Wilson2019}, we determine the region within the primary's envelope where convection transports liberated orbital energy to the surface faster then the orbit decays.  In this regime, the CE self-regulates, convectively cools, and qualitatively differs from non-convective CEs as the orbital energy is lost from the system.  From this, we construct light curves for convective common envelopes, and compare these results to the upcoming observatory's specifications. In particular, this work found that:

\begin{enumerate}
    \item Light curves for CE targets change when convection is considered. Including convection increases the duration of the observable light curve, with the more massive primaries being more strongly affected by convective effects than the less massive primaries (see Figure~\ref{fig:lightcurves}).
    \item The Simonyi Survey Telescope can detect these events, not only in our galaxy but throughout the Local Group as well. This will establish observational constraints that can help distinguish what role convection has in CE evolution (see Figures \ref{fig:appmags}, \ref{fig:howlongall}).
\end{enumerate}

These findings motivate using LSST data to identify common envelope transient events and suggest a likelihood of observation. Due to its modular structure, the method described in this work can be applied to current instruments as well, as it is possible CE events could be identifiable within archival data.

The models produced here were pairings of low-mass stars and companions with well-defined parameters that would unbind the primary star's envelope while allowing the convective motions within the CE to remain subsonic (Figure \ref{fig:draglums}). Future work specific to the observatory includes producing models for other pairings that would be visible given the telescope's parameters, such that a wider range of CE events can be identified once Rubin comes online. Because this same process can be applied to currently available data, confirmation of convective CE events can be done with any instrument sensitive to these parameters.

Further work also includes understanding the effects of convection on light curves of massive stars. This would improve our understanding of CE evolution overall, as stars with masses greater than those described in this work have convective regions that do not penetrate as deeply into the primary's envelope \citep{Wilson2022}. The discrepancy in how efficiently orbital energy can be tapped to drive ejection between low-mass and high-mass CEs has already had success explaining post-CE population distributions.  Identifying objects undergoing such physical processes in real time would provide useful observational constraints on CE physics.

CE evolution is anticipated to occur on timescales well within the scope of the Rubin observatory, which will observe the southern sky every few days over a decade in the visible light spectrum \citep{LSST2019}. While many post-CE objects have been, or are being, identified \citep{Grondin2024}, targets currently undergoing CE evolution have yet to be recognized due to the short duration of the CE phase and the obscuring envelope of the primary. The findings outlined in this work help overcome these difficulties, and can aid in identification, regardless of the instrument.

\section*{Acknowledgements}
The authors acknowledge Maria Drout, Phil Muirhead, Steffani Grondin, Josh Faber, Joel Kastner, Jared Goldberg, Matteo Cantiello, and Carlos Badenas for helpful conversation and comments regarding this work. NN is supported by the New York Space Grant Consortium.  JN acknowledges support from NSF AST-2009713 and AST-2319326.

\section{Data Availability}
No new observational data were generated from this research. Data underlying this article is available in the articles and supplementary materials of the referenced papers. Stellar interior models were derived from MESA, which is available at http://mesa.sourceforge.net. Filter values for anticipated magnitudes were calculated using resources found on the LSST website, as well as specifications described in \citealt{Kahn2010}. Additional information is available by request.

\bibliographystyle{elsarticle-harv}

\bibliography{references}

\end{document}